\documentclass[twocolumn,showpacs,superscriptaddress,preprintnumbers,floatfix]{revtex4}
\usepackage{graphicx}

\def\be{\begin{equation}}
\def\ee{\end{equation}}
\def\bea{\begin{eqnarray}}
\def\eea{\end{eqnarray}}

\begin{document}

\preprint{astro-ph/yymmddd}

\title{Constraining Born-Infeld models of dark energy with CMB 
anisotropies}

\author{L. Raul Abramo}
\email{abramo@fma.if.usp.br}
\affiliation{Instituto de F\'{\i}sica, 
Universidade de S\~ao Paulo \\
CP 66318, 05315-970 S\~ao Paulo, Brazil}

\author{Fabio Finelli}
\email{finelli@bo.iasf.cnr.it}
\altaffiliation[Also supported by ]{ INFN, Sezione di Bologna, 
via Irnerio 46 -- I-40126 Bologna -- Italy}
\affiliation{IASF/CNR, Istituto di Astrofisica Spaziale e
Fisica Cosmica, Consiglio Nazionale delle Ricerche \\
via Gobetti, 101 -- I-40129 Bologna -- Italy}

\author{Thiago S. Pereira}
\email{thiago@fma.if.usp.br}
\affiliation{Instituto de F\'{\i}sica, 
Universidade de S\~ao Paulo \\
CP 66318, 05315-970 S\~ao Paulo, Brazil}

\date{\today} 

\begin{abstract} 
We study the CMB constraints on two Dark Energy models described by scalar 
fields with different Lagrangians, namely a Klein-Gordon and a 
Born-Infeld field. The speed of sound of field fluctuations are 
different in these two theories, and therefore the predictions for CMB and
structure formation are different.
Employing the WMAP data on CMB,
we make a likelihood analysis on a grid of theoretical models. We
constrain the parameters of the models and compute the probability
distribution functions for the equation of state. We show 
that the effect of the different sound speeds affects the low multipoles 
of CMB anisotropies, but is at most marginal for the class of models 
studied here.
\end{abstract}


\maketitle

\section*{Introduction}

Several observations now indicate \cite{WMAP,SN} that
the universe has been accelerating its expansion rate for the 
last 5 -- 10 Gy.
Standard candidates for the ``dark energy'' (DE) which is
responsible for this recent burst of expansion are a
cosmological constant and a scalar field 
\cite{RP,accellera,kessence,japan,AF03} 
which, for reasons unknown yet, started to dominate over the other 
types of 
matter just at our cosmological era. Unfortunately, DE seems to
pose an even more formidable problem than that of dark matter to 
observational cosmology: whereas we still hope that the particles that 
make up dark matter will eventually be detected, and that
incremental knowledge about the distribution of large-scale sctructure
and galactic dynamics will eventually nail down the basic properties of 
dark matter and its relation to baryonic matter, no such hopes apply for 
DE, at least for now. Nevertheless, we must not shy away from 
trying to test models of DE, under penalty of theory running 
amok.

One of the best opportunities to test DE
is by looking at the anisotropies in the temperature and polarization 
of the 
cosmic microwave background radiation (CMB), which have been measured
with exquisite accuracy by WMAP \cite{WMAP} --- and 
will be measured with 
even grander precision by the PLANCK mission \cite{PLANCK}.

There are basically two ways through which DE can affect the 
CMB. The first is through its impact on the expansion rate, which 
is determined by the equation of state of the DE component.
The second is directly through the perturbations: 
DE, if it is not a plain cosmological constant, possesses 
small inhomogeneities which interact gravitationally with the 
inhomogeneities in baryons, dark matter 
and relativistic matter. The physical properties of DE perturbations
constitute additional ingredients which can impact the CMB 
anisotropies and LSS.

Here we are interested in the role of the sound speed of DE 
perturbations. Whereas for the background evolution it
is only necessary to specify the DE budget at the present time 
$\Omega_{X \, 0}$, and its pressure/density ratio $w_X (t) = p_X (t) / 
\rho_X (t)$, in order to 
classify the perturbations sector we need to specify the pressure 
perturbations as well \cite{erickson}:
\be
\delta p_{X} = c_{X}^2 \delta \rho_{X} + 3 H (1+w_{X}) 
\frac{\theta_{X} \rho_{X}}{k^2} \left( c_{X}^2 -
\frac{\dot p_{X}}{\dot \rho_{X}} \right) \,,
\label{deltap}
\ee
where $\rho_{X}$, $\delta \rho_{X}$, $c_{X}^2$ and $\theta_{X}$
are respectively the DE density, density perturbation, sound speed and 
velocity potential. The case for a perfect fluid DE model in which $\delta
p_X = c_{X}^2 \delta \rho_{X}$ has been analyzed in
\cite{CF} and compared with the WMAP data in \cite{AFBC}. In this paper we 
focus on the scalar field case in which the pressure perturbation is not 
simply related to the density perturbation.

The standard {\it quintessence} scenario (a canonical scalar field 
described by a Klein-Gordon Lagrangian) is a minimal 
modification to $\Lambda$CDM, and in that particular case
$c_X^2=1$. All the information contained in the 
scalar potential $V(\varphi)$ is encoded in $w_X$ in that case.

The possibility of having $c_X^2 \ne 1$ in the context of 
scalar field theories by considering a
Lagrangian with a non-standard kinetic term is not new.
If such non-canonical scalar field plays the role of an 
inflaton, the predictions for scalar and tensor perturbations
are different with respect to the usual (canonical) scenario 
\cite{kinflation}. In the context of DE, K-essence 
\cite{kessence} was proposed (see also \cite{japan}), even 
if it has the unpleasant feature of having fluctuations which
can travel with speeds faster than that of light. The imprints 
of K-essence on CMB anisotropies have already been studied elsewhere
\cite{erickson,kessence_pheno}, but not yet in a statistic way.
Here we focus on a different scalar field theory and we 
perform a statistical study of the predictions \footnote{
See also
\cite{beandore} for a completely phenomenological approach with $w_X$
constant in time.}.

To be concrete we study a Klein-Gordon (KG henceforth) Lagrangian density:
\be
\label{L_csf}
{\cal{L}}_{KG} = 
- \sqrt{-g} \, 
\left[ \frac12 \partial^\mu \varphi
\partial_\mu \varphi + V(\varphi) \right] \; ,
\ee
and compare it to a Born-Infeld \cite{borninfeld} (BI henceforth) 
Lagrangian 
\be
\label{L_BI}
{\cal{L}}_{BI} = - V (\phi) \sqrt{1+
\frac{\partial^\mu \phi \partial_\mu 
\phi}{M^4}} 
\ee
(where we have assumed a signature $(-+++)$ for the metric and we have 
introduced the mass scale $M$).
This Lagrangian can be thought as a field theory generalization 
of the Lagrangian of a relativistic particle \cite{padma}.
For $\partial^\mu \phi \partial_\mu \phi/M^4 << 1$, by taking the first 
term of a Taylor expansion of Eq. (\ref{L_BI}), we get a Lagrangian which 
is equivalent to the KG one by redefining the field (except for 
problems with the invertibility of this redefinition). Only by taking 
the full Lagrangian (\ref{L_BI}), the two theories are different. The 
BI was recently revisited in connection 
with string theory, since it seems to represent a low-energy effective
theory of D-branes and open strings \cite{Sen,gg}.
In the case of constant potential, the model reduces to a
type of hydrodynamical matter known as the Chaplygin Gas \cite{Chap} ---
whose predictions were studied in \cite{CF,AFBC}.
While for the KG Lagrangian the sound speed of 
fluctuations is 
$1$ independently of the potential, in the BI case 
$c_X^2 = -  w_X \geq 0$ always, regardless of the potential. 

For both theories we will assume an inverse power-law potential:
\be
\label{pot_T}
V(f) = \frac{ m^{4+p}}{f^{p}} \; ,
\ee
where $p>0$ and $f$ stands for $\varphi$ or $\phi$ in the canonical or
non-canonical cases, respectively. 
Positive powers of the field
($p<0$) yield models which, if not finely tuned, do not lead to 
acceleration. In the following we split the fields as $f = f_0 (t) + 
\delta f (t, {\bf x})$, as usually done in cosmology.

\section*{Equation of state versus sound speed}

In the canonical case, the scalar field model with potential (\ref{pot_T})
is known as the Ratra-Peebles model of dark energy \cite{RP}.
In the BI case it was shown in \cite{AF03} that this
model can lead to an accelerating regime when $p \leq 2$.
With the potential given by Eq. (\ref{pot_T}), both theories behave
very similarly at the level of the background. This is so because
both scalar fields possess fluid-like attractor solutions
when the background energy density is dominated by a perfect
fluid such as dust ($w_F=0$) or radiation ($w_F=1/3$). 
For the Ratra-Peebles model the equation of state of the attractor
during a fluid-dominated period is \cite{ZWS}:
\be
\label{es_c}
w_\varphi = \frac{\frac12 \dot\varphi_0^2 - V(\varphi_0)}
{\frac12 \dot\varphi_0^2 + V(\varphi_0)} \simeq \frac{p \, 
w_F - 2}{p+2} \; ,
\ee
where $\varphi_0(t)$ is the homogeneous value of the scalar field.
For the BI model the equation of state during a fluid-dominated
period is \cite{AF03}:
\be
\label{es_T}
w_\phi = -1 + \frac{\dot\phi_0^2}{M^4} \simeq -1 + p 
\frac{1 + w_F}{2} \; .
\ee
In particular, for $p \ll 1$, $w_\varphi \simeq w_\phi$, while tracking is 
more and more accurate for $p >>1$ in the KG case and for $p \simeq 2$ in 
the BI case. 

\begin{figure}
\includegraphics[width=8cm]{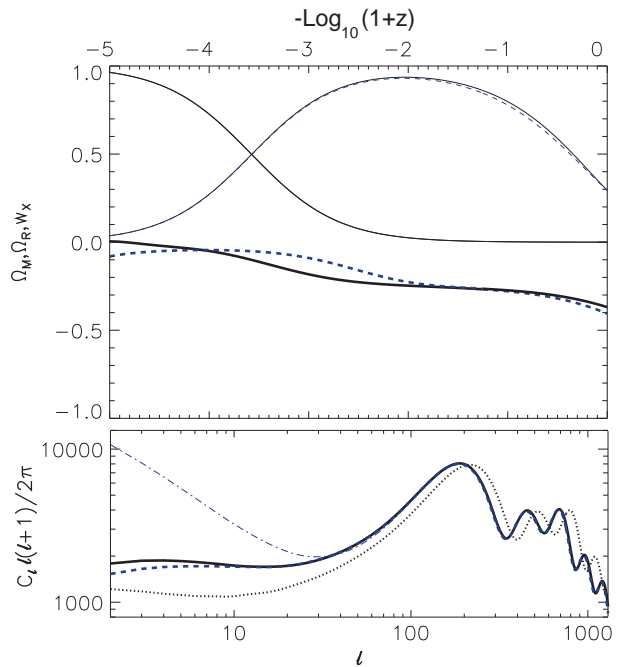}
\caption{\label{fig:1}
Upper panel: background quantities 
for two similar Born-Infeld (BI, dashed lines) and Klein-Gordon 
(KG, solid lines)
dark energy models as a function of redshift. The upper curves
denote $\Omega_r$ and $\Omega_m$, and the lower, nearly superimposed 
curves denote the equations of state for the two models.
The cosmological parameters are 
$\Omega_b=0.04$, $\Omega_m=0.26$, $h=0.72$, and either
$p=3/2$ (BI) or $p=6$ (KG.)
Lower panel: temperature angular power spectra
for KG model (solid line), 
BI model (dashed line) and a fiducial $\Lambda$CDM (dotted line), 
all normalized at the location of their first doppler peak. Only for 
comparison we also show the (wrong) power spectrum of the BI model 
obtained without considering the scalar field perturbations (dot-dashed line).}
\end{figure}

The main difference between them lies in the behaviour of their
perturbations. The canonical scalar field perturbations 
$\delta\varphi(x,t)$ obey the equation of motion:
\be
\label{eom_c}
\delta \ddot \varphi - \frac{\nabla^2}{a^2} \delta\varphi
+ 3 H \delta \dot\varphi +  V_{,\varphi \varphi} \delta\varphi 
= -\frac12 \dot{h} \dot\varphi_0 \; ,
\ee
where $a$ is the scale factor and
$h$ is the synchronous gauge metric perturbation. 
The BI scalar
perturbation, on the other hand, obeys the equation:
\begin{eqnarray}
\nonumber
\frac{\delta \ddot\phi}{1 - \dot \phi_0^2/M^4} 
&-& \frac{\nabla^2}{a^2} \delta\phi + 3 H \delta \dot\phi
+ \frac{2 \dot \phi_0 \ddot \phi_0}{(1 - \dot \phi_0^2/M^4)^2} 
\delta \dot\phi \\ 
\label{eom_T}
&+& M^4 (\log V)_{,\phi \phi} \delta\phi = - \frac12 \dot{h} \dot\phi_0 
\,.
\end{eqnarray}
Eqs. of motion (\ref{eom_c}) and (\ref{eom_T})
determine how perturbations behave:
while the canonical scalar field perturbations have
a speed of sound $c_{\varphi}^2=1$, in the BI case the speed of 
sound is $c_\phi^2 = - w_\phi = 1-\dot\phi_0^2/M^4$.
Since $c_X^2$ determines how fast fluctuations dissipate, 
a lower sound speed increases the phase space of modes
which are Jeans unstable \cite{beandore} [to make 
contact with Eq. (\ref{deltap}), $\theta_X = k^2 \delta f/\dot f$, 
where $\delta f$ is the field perturbation.]
Hence, the spectra of anisotropies predicted by the two 
models are slightly different because of the different Jeans scales in 
the DE sector -- see Fig. 1.

A key feature of these models is that not only their backgrounds are
well described by a simple attractor: their perturbations are
funnelled to an attractor as well \cite{AF01}. This means that the
issue of initial conditions for the background field and the perturbations 
is partially solved \footnote{We have checked this numerically,
over several orders of magnitude, both for the background scalar field 
and for the perturbations}. If this was not true, 
initial conditions would constitute additional free parameters in 
the models.

\section*{CMB phenomenology and likelihood analysis}

In order to compute the CMB anisotropies we employed a modified version
of the code CMBFAST v. 4.1 \cite{CMBFAST}, adding scalar fields to the
system of equations.

We take the following vector of free parameters:
\be
\label{free_param}
\vec{\alpha} = \{ H, \Omega_m, \Omega_b, n_s , m, p ; A \} \; ,
\ee
where $H$ denotes the present value of the Hubble constant in Km
s$^{-1}$ Mpc$^{-1}$, $\Omega_m$ is the amount of dark matter,
$\Omega_b$ the amount of baryons, $n_s$ is the
scalar spectral index,
$m$ is the scalar field mass 
scale in the potential (\ref{pot_T}) 
and $p$ is the power of the scalar potential. As always, there is a 
free parameter ({\it A}) 
related to the (unobservable) overall normalization of 
the CMB spectrum.
Since we consider only flat models, 
the density in the dark energy component, $\Omega_{X}$, is 
fixed once the other cosmological parameters are determined.
We have ommitted the BI scale $M$ from the space of parameters since
it can be absorbed by a redefinition of $m$.
\begin{figure}
\includegraphics[height=6cm]{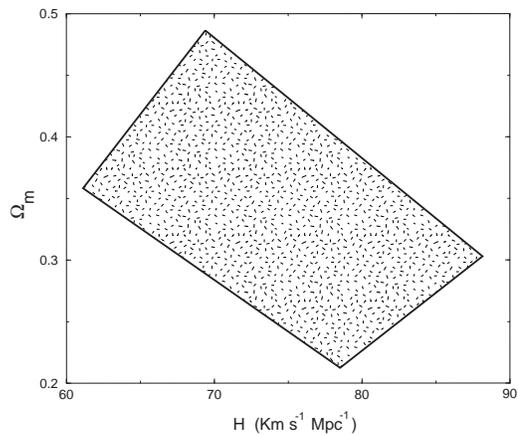}
\caption{\label{fig:2}
Region of $[H,\Omega_m]$ probed by our simulations.} 
\end{figure}


\vskip 0.3cm

We have evaluated the likelihood function $\mathcal{L}(C_l|\vec\alpha)$ 
for a grid of roughly 370,000 models using the code described in
\cite{verde}. In order to get the posterior probability distribution function
(p.d.f) for a given parameter, we must marginalize the likelihood
function over the remaining ones. Since $\mathcal{L}(C_l|\vec\alpha)$ 
is not a p.d.f in the usual sense, we use Bayes' Theorem:

\begin{equation}
\label{pdf}
P(\alpha_i) \propto \int \mathcal{L}(C_l|\vec{\alpha})
\prod_{j\neq i} \mathcal{P}(\alpha_j) d \alpha_j \; ,
\end{equation} 
where $\mathcal{P}(\alpha_j)$ 
is the prior p.d.f for $\alpha_j$. 
\begin{figure}
\includegraphics[width=6cm]{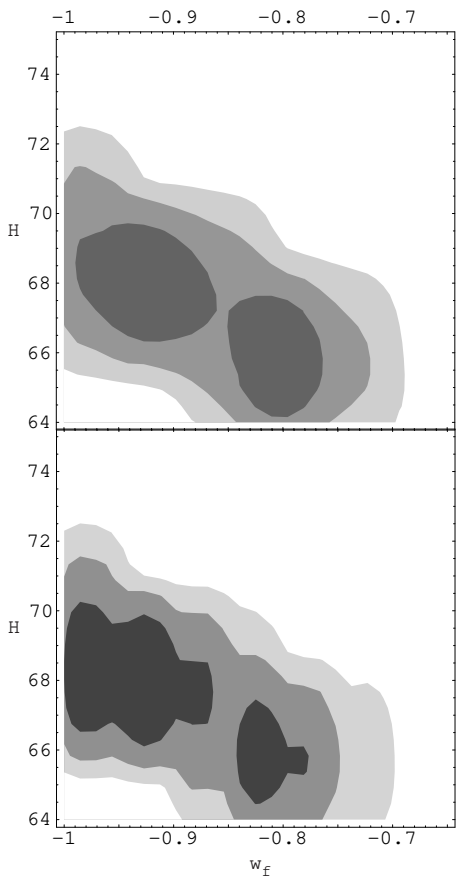}
\caption{\label{fig:3}
Joint posteriori p.d.f. for $H$ and $\omega_f$ from WMAP data,
for Ratra-Peebles (upper panel) and Born-Infeld (lower panel) 
models. The shaded areas correspond to 1, 2 and 3$\sigma$ levels.}
\end{figure}

For the parameters in (\ref{free_param}) we set an
equally spaced grid with the following (flat) priors: 64 $\leq
H \leq$ 80 , 0.036 $\leq \Omega_b \leq$ 0.069, 0.95 $\leq n_s
\leq$ 1.09. In the matter era ($\omega_F=0$), 
the power $p$ of the potential can be substituted
by $\omega_f$ because of Eqs. (\ref{es_c})-(\ref{es_T}), 
so we chose a grid that was equally spaced in $\omega_f$ rather than $p$, with
$ -0.6 \leq \omega_f \leq -1$. Regarding $\Omega_m$ and $H$, 
it was numerically more suitable to use a certain combination of the 
two parameters which effectively explored a region of parameter space 
whose shape is shown in Fig. 2.

We can study the parameters $H$ and $\omega_f$ (or, 
equivalently, $p$) jointly by marginalizing against (i.e., 
integrating out) all other parameters
according to (\ref{pdf}), thus obtaining the {\it posteriori} 
marginalized p.d.f. $P_{m}(H,\omega_f)$.
This joint p.d.f for $H$ and $\omega_f$ is presented
in Fig. 3 for the models under scrutiny.
The contours correspond to 1, 2 and $3\sigma$ levels.
For the Ratra-Peebles (KG) case, the joint p.d.f for $H$ and
$\omega_f$ agrees with \cite{CD}.

We are mainly interested in $\omega_f$, so we marginalize
against $H$ as well to obtain the
{\it posteriori} p.d.f. for the equation of state
of dark energy.
The p.d.f's, shown in Fig. 4, have been smoothed through a 
Bezier spline so that the effects of having an imperfect
grid are not overestimated.
\begin{figure}
\includegraphics[width=8cm]{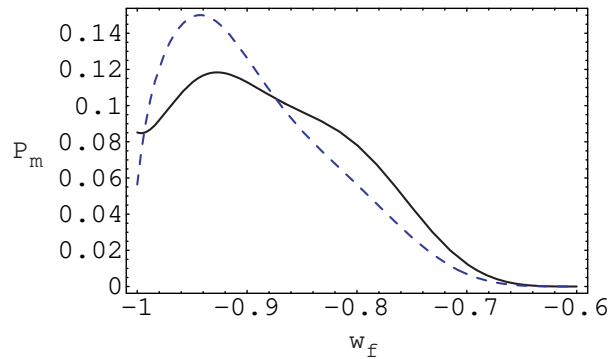}
\caption{\label{fig:4}
Probability distribution function as a function of the equation of
state for both models: KG (solid line) and BI (dashed line.)}
\end{figure}

The limits on the equation of state of the dark energy component 
are, for the KG model:
\be
\nonumber
\omega_\varphi > -0.84 \, (1\sigma), \, -0.74 \, (2 \sigma), \, -0.68
\, (3\sigma) \, ,
\ee
where 1$\sigma$, 2$\sigma$ and 3$\sigma$ correspond to 68.3\% 
C.L., 95.4\% and 99.7\% C.L., respectively. 
For the BI model the limits on its equation of state are:
\be
\label{limw_BI}
\omega_\phi > -0.87 (1\sigma), \, -0.76 \, (2\sigma), \, -0.69 \,
(3\sigma) \, .
\ee


\section*{Conclusions}

We have tested two scalar field models of dark energy against the 
WMAP data on CMB anisotropies. For simplicity we have taken the same 
potential for both Lagrangians. Our aim was to
compare models which had nearly identical backgrounds, but
whose perturbative sectors behave differently. We did this by
comparing two scalar field models, a Klein-Gordon (the usual
quintessence)
and a Born-Infeld scalar,
both of which have very similar attractor solutions during 
radiation- and matter-domination.

From a fundamental perspective, we have shown that a BI scalar field can 
easily play the role of DE. As for being a CDM candidate 
\cite{CDM} the 
non-linear stage must still be carefully analyzed \cite{nonlinear}.
We stress that a very interesting feature of BI theories is that
the BI scalar can act as dust and drive the universe 
into acceleration, where the trigger to the 
accelerated phase is the transition of the background from 
radiation-dominated to dust-dominated at $z_{eq} \sim 10^4$ \cite{AF03}.

For the Ratra-Peebles model (a canonical scalar field model) we find an
allowed range of parameters $p < 0.38$ at 1$\sigma$, corresponding
to an equation of state during the attractor regime in the matter-dominated
period of $-1 \leq w_\varphi \leq -0.84$. The Born-Infeld model,
on the other hand, is more tightly constrained: we find $p<0.27$
at 1$\sigma$, corresponding to an equation of state of the attractor
regime of $-1 \leq w_\phi \leq -0.87$.
For the Born-Infeld model, this also implies a limit for the sound 
speed of dark energy $c_\phi^2>0.87$ at the $1 \sigma$ level.

The effect of the perturbations was not 
dramatic in the BI case, since we have $c_X^2 
= -w_X$ (compared to $c_X^2 =1$ for quintessence). 
However, in the cases studied here, the speed of sound 
affects interestingly the low-$\ell$ multipoles.
In order to have a DE model which fits observations well, one has to 
consider $w_X \lesssim -0.5$ and therefore the sound speed for BI 
fluctuations is not 
sufficiently different from quintessence. Because of cosmic variance and 
high error bars, low $\ell$'s have a diminished influence on the 
Likelihood, which
means that the region where the different perturbative behaviors
show up more conspicuously are not well represented in the Likelihood 
analysis. Nevertheless, changing the
speed of sound of dark energy by a factor of order 10\% can have an
impact (from the point of view of the Likelihoods) of the same order
of magnitude as changing the other cosmological parameters by a few 
percent, which is the precision target of future experiments such as 
PLANCK \cite{PLANCK}.

This reinforces the notion \cite{kessence_pheno} 
that only dark energy models which possess a highly unusual perturbative
behavior, such as certain models of k-essence \cite{kessence}, models 
with $c_X^2 \sim 0$ (and unrelated to $w_X$) or dark energy perfect 
fluids 
\cite{CF,AFBC} (where the pressure perturbation in Eq. (\ref{deltap}) 
depends only on the density perturbation)
can have a large impact on the CMB.

\vskip 0.5cm

\noindent
We would like to thank V. Mukhanov and I. Waga for useful 
conversations.
F. F. would like to thank the Instituto de F\'{\i}sica, 
Universidade de S\~ao Paulo, for its warm hospitality when
this work was initiated. R. A. would like to thank the CNR/IASF and INFN - 
Bologna as well, for its hospitality. This work was also supported by 
FAPESP, CNPq and CAPES.



\end{document}